\documentclass[aps,prb,twocolumn,showpacs,superscriptaddress,floatfix]{revtex4-2}
\usepackage{graphicx}
\usepackage{amsmath}
\usepackage{amssymb}
\usepackage{bm}
\usepackage{hyperref}
\usepackage{cleveref}

\usepackage{array}

\usepackage{lipsum} 

\begin{document}
\title{Advancing Magnetic Materials Discovery - A structure-based machine learning approach for magnetic ordering and magnetic moment prediction}
\author{
Apoorv Verma$^{1,*}$,
Junaid Jami$^{1,*}$,
Amrita Bhattacharya$^1$ \\
\small $^1$AbCMS Lab, Department of Metallurgical Engineering and Materials Science, \\
\small Indian Institute of Technology Bombay, Mumbai, Maharashtra 400076, India \\
\small $^*$These authors contributed equally to this work.
}
\date{\today}

\begin{abstract}

Accurately predicting magnetic behavior across diverse materials systems remains a longstanding challenge due to the complex interplay of structural and electronic factors and is pivotal for the accelerated discovery and design of next-generation magnetic materials. In this work, a refined descriptor is proposed that significantly improves the prediction of two critical magnetic properties - magnetic ordering (Ferromagnetic vs Ferrimagnetic) and magnetic moment per atom - using only the structural information of materials. Unlike previous models limited to Mn-based or lanthanide-transition metal compounds, the present approach generalizes across a diverse dataset of 5741 stable, binary and ternary, ferromagnetic and ferrimagnetic compounds sourced from the Materials Project. Leveraging an enriched elemental vector representation and advanced feature engineering, including nonlinear terms and reduced matrix sparsity, the LightGBM-based model achieves an accuracy of 82.4\% for magnetic ordering classification and balanced recall across FM and FiM classes, addressing a key limitation in prior studies. The model predicts magnetic moment per atom with a correlation coefficient of 0.93, surpassing the Hund’s matrix and Orbital field matrix descriptors. Additionally, it accurately estimates formation energy per atom, enabling assessment of both magnetic behavior and material stability. This generalized and computationally efficient framework offers a robust tool for high-throughput screening of magnetic materials with tailored properties.
\end{abstract}
\maketitle

\section{Introduction}

Advancements across a broad spectrum of modern technologies - including sustainable energy systems~\citep{magnetsgreenenergy}, data storage~\citep{magneticstoragedevice1,magneticstoragedevice2,magneticstoragedevice3}, advanced transportation~\citep{magnetstrains}, and emerging applications such as magnetic refrigeration, quantum computing, and medical imaging~\citep{magnetsrefrigeration,magnetsquantumcomputers,magnetsimaging,application4,application5} - are fundamentally enabled by high-performance magnetic materials. As the demand for next-generation technologies intensifies across multiple sectors, the discovery of novel magnetic materials with enhanced functional properties and cost-effectiveness is essential not only to drive innovation but also to ensure scalability and economic viability. Consequently, the exploration and design of advanced magnetic materials remains a vital area of research across both established and emerging applications. In recent decades, numerous computational tools have emerged to simulate magnetic phenomena~\citep{quantumespresso, OOMMF, spirit, vampire, wien2k}, providing an efficient alternative to experimental techniques like neutron scattering \citep{experiment_neutron} and resonant X-ray scattering \citep{experiment_xray}, which, though offer atomic-scale characterization of magnetic structures, are constrained by high costs, limited access, and time-intensive processes. First-principles methods, particularly density functional theory (DFT), have been extensively employed to calculate key magnetic properties-such as magneto-crystalline anisotropy~\citep{dftanisotropy}, magnetization~\citep{dftmagneticmoment}, and exchange interactions~\citep{dftexchange}-which together govern a material’s magnetic behavior and technological applicability. However, as system complexity increases due to larger atomic configurations and more intricate interactions, the computational cost of such \textit{ab initio} methods grows substantially. 

Considerable efforts have also been directed towards the accurate prediction of magnetic ordering and traditionally, this has involved exhaustive DFT calculations across all plausible magnetic configurations \citep{dft_ordering2} - ferromagnetic (FM), antiferromagnetic (AFM), and ferrimagnetic (FiM). While effective, this brute-force method becomes computationally very expensive, especially for materials with large unit cells or multiple magnetic sublattices. To address these challenges, several strategies have emerged to develop algorithms capable of pre-screening magnetic configurations and establishing a priority order, thereby narrowing down the search space before applying computationally expensive methods \citep{dft_filter1,dft_filter2,article,ZHENG2021107659}. Horton \textit{et al.}~\citep{article} proposed a robust and automated high-throughput DFT+U workflow for determining the magnetic ground state of inorganic materials, systematically enumerating and prioritizing collinear magnetic orderings based on symmetry. Their workflow successfully predicts the experimental ground state in over 60\% of the materials benchmarked. Similarly, MagGene \citep{ZHENG2021107659} leverages a genetic algorithm to efficiently explore the space of possible magnetic configuration space, iteratively evolving promising candidates rather than exhaustively evaluating all possibilities, making the search for magnetic ground states far more efficient-even in systems with complex collinear and non-collinear spin arrangements.

\begin{figure*}
    \centering
    \includegraphics[width=\linewidth]{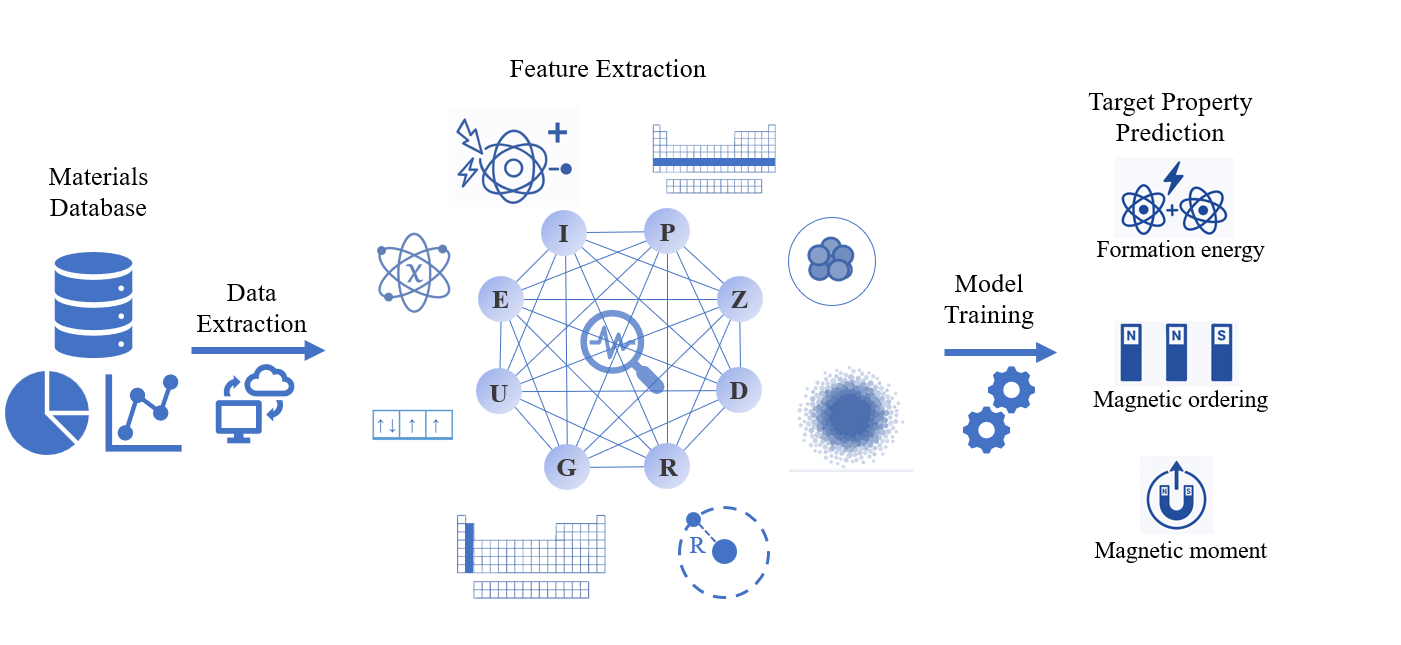}
    \caption{Workflow for predicting material properties using ML. Structural data is extracted from a materials database, followed by feature extraction to obtain numerical descriptors, which are then used to train an ML model for predicting target properties}
    \label{fig:descriptor} 
\end{figure*}

While these advances in DFT workflows and search algorithms have significantly improved the efficiency of predicting magnetic properties, they still rely on repeated DFT calculations. This limits their scalability, particularly in high-throughput screening - motivating the integration of data-driven machine learning (ML) approaches that can further accelerate and scale magnetic materials discovery. Leveraging large-scale materials databases-such as the Materials Project~\citep{Jain2020TheMP}, AFLOW~\citep{CURTAROLO2012218}, NOMAD~\citep{Draxl2018NOMADTF}, OQMD~\citep{5bec5ceb51a04553b49867290602a0f4}, and Novamag~\citep{novamag} - ML can help predict material properties without the heavy computational load typically associated with \textit{ab-initio} methods. By leveraging readily accessible features like atomic composition and crystal structure, and techniques such as neural networks, support vector machines, and decision trees, these models facilitate rapid screening of material properties~\citep{MAL2024171590,D1TC03776E,b20alloys,PhysRevMaterials.6.024402}. ML-driven studies have successfully identified rare-earth-free permanent magnets with enhanced magneto-crystalline anisotropy and magnetization~\citep{MAL2024171590,D1TC03776E}, enhanced transition temperatures in B20 chiral magnets for spintronic applications~\citep{b20alloys}, and resolved site-specific anisotropy in Fe-Co nitrides~\citep{PhysRevMaterials.6.024402}. 

ML techniques have also gained traction as powerful tools for magnetic ordering classification. For instance, a study on 366 uranium-based compounds \citep{PhysRevMaterials.8.114405} achieved a classification accuracy of 62.1\% across FM, AFM, and paramagnetic categories - although it did not consider FiM states. Merker \textit{et al.} \citep{momlc2} employed euclidean neural networks to classify magnetic ordering directly from atomic coordinates of materials containing transition metals and rare earth elements. Their model achieved 77.8\% accuracy for the ternary classification scheme (FM/FiM, AFM, and non-magnetic), but grouped FM and FiM into a single class based on non-zero net magnetization. Frey \textit{et al.}~\citep{momlc3} conducted a large-scale high-throughput DFT study on over 3,000 transition metal oxides, generating more than 27,000 unique magnetic configurations. Their ML classifier is trained to predict magnetic ordering, but again did not treat FiM states distinctly. Instead, FiM configurations  were labeled as FM or AFM based on whether the net magnetization exceeded a threshold of $0.1~\mu_{\mathrm{B}}$ per unit cell - an oversimplification that fails to account for the distinct sublattice magnetization profiles of FiM systems. Although such approaches report promising overall classification accuracies, a critical examination reveals a recurring blind spot - the improper handling or complete neglect of FiM states. This systematic misclassification or omission of ferrimagnetism not only affects model performance but also hinders physical interpretability. Given the technological importance of ferrimagnetic materials in magneto-optical systems \citep{fim-magneto-optical}, passive microwave components such as isolators, circulators and phase shifters \citep{fim-microwaveferrites} and particularly for emerging emerging high-density, high-speed and high-efficiency spintronic applications \citep{fim1,fim2,fim3,fim4,fim5}, addressing this oversight is crucial for advancing predictive magnetic materials design.

\begin{figure}[t!] 
    \centering
    \includegraphics[width=\linewidth]{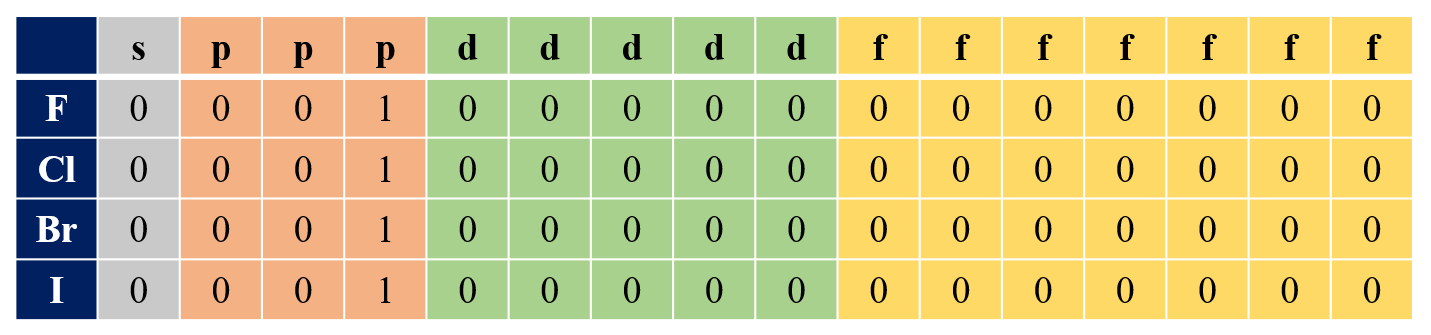}
    \caption{In the Hund’s matrix method, elements like fluorine (\([ \text{He} ] 2s^2 2p^5\)), chlorine (\([ \text{Ne} ] 3s^2 3p^5\)), bromine (\([ \text{Ar} ] 3d^{10} 4s^2 4p^5\)), and iodine (\([ \text{Kr} ] 4d^{10} 5s^2 5p^5\)) - despite being chemically distinct-are represented identically due to their similar valence shell electronic configurations. This limitation arises because the method primarily encodes information based on valence electron configuration, which fails to capture the nuanced differences in chemical behavior and atomic size among these elements.}
    \label{fig:hunds drawback} 
\end{figure}

To address this, a descriptor known as Hund’s matrix \citep{KHATRI2024172026}, has been recently proposed which demonstrated improved performance in predicting magnetic properties - specifically magnetic ordering and magnetic moment per atom of Mn-based materials. It captures the essential magnetic characteristics of a material by encoding the valence shell electronic configuration and interactions of its unpaired valence electrons. Each atom is represented as a $1 \times 16$ vector, corresponding to the occupancy of the \textit{s}, \textit{p}, \textit{d}, and \textit{f} orbitals, with 1 denoting unpaired electrons and 0 for paired electrons. The matrix representation of a material is then constructed using -

\begin{equation}
\text{HM} = \sum_{i=1}^{n} \sum_{j=1}^{nn_i} \frac{\mathbf{v}_i^\top \cdot \mathbf{v}_j}{d_{ij}^2}
\end{equation}

where $n$ is the number of atoms in the unit cell, $nn_i$ is the coordination number of atom $i$, $\mathbf{v}_i$ is the vector for atom $i$, and $d_{ij}$ is the distance between atoms $i$ and $j$. However, this method has notable limitations. Consider elements from Group 17 - fluorine, chlorine, bromine and iodine - despite being chemically distinct, are represented identically in Hund’s matrix method, as shown in Fig.~\ref{fig:hunds drawback}, because they share a similar valence shell electronic configuration. Although the incorporation of interatomic distances enables the descriptor to differentiate between local atomic environments, it does not resolve the underlying issue of elemental indistinguishability. Consequently, critical atomic-level features such as atomic number, atomic radius, electronegativity, and ionization energy - factors that can significantly influence a material's overall properties - are entirely omitted, limiting the descriptor’s capacity to capture nuanced structure-property relationships.

To overcome these limitations, the present work proposes a refined approach that constructs a more comprehensive and information-rich descriptor for elemental representation. The proposed descriptor is systematically evaluated against established structural representations, with a focus on its performance in predicting key material properties - magnetic ordering, magnetic moment per atom and formation energy. It is designed to work across different elements and crystal structures without any restrictions, and is applied uniformly across all stable, binary and ternary, FM and FiM compounds present in the Materials Project database~\citep{Jain2020TheMP}. This broad applicability allows the model to capture a wider spectrum of chemical environments, thereby enhancing its generalizability and robustness in materials prediction tasks.

\section{Methodology}

\subsection{Data Collection}
\begin{figure}[t!] 
    \centering
    \includegraphics[width=0.9\linewidth]{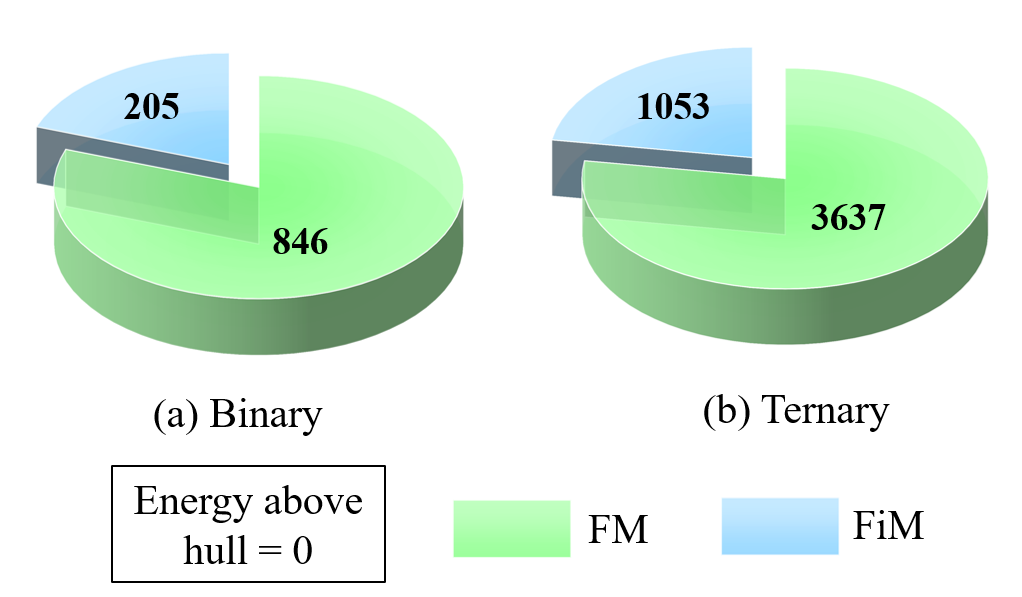}
    \caption{Distribution of FM and FiM materials in the dataset (a) Binary compounds consist of 846 FM and 205 FiM materials (b) Ternary compounds include 3,637 FM and 1,053 FiM materials}
    \label{fig:dataset composition} 
\end{figure}

The materials data used in this work has been sourced from the Materials Project database \citep{Jain2020TheMP}, which provides access to a vast collection of over 178,000 computationally and experimentally derived materials data entries, supporting researchers in the discovery and optimization of materials. All stable, binary and ternary, FM and FiM compounds available on the database (totaling to 5741) have been used for the present work. Compounds with an energy above hull equal to zero have been considered stable. Notably, no specific elemental filter or symmetry criteria were applied in the compound selection process from the database, resulting in a highly diverse and complex dataset, which presents a challenging task for mapping magnetic properties. AFM compounds have been excluded from the training dataset for two key reasons - first, they represent a significantly smaller subset of compounds leading to a severe class imbalance that would hinder effective training of a multi-class machine learning model and second, the central objective of this study is to disentangle and accurately classify FM and FiM materials, a challenge that has been largely overlooked in previous works as FiM systems have often been misclassified, grouped with FM, or entirely omitted. By isolating these two magnetically ordered categories, the aim is to fill a critical gap in data-driven magnetic materials research and support more precise modeling. Fig. ~\ref{fig:dataset composition} presents the distribution of magnetic ordering across binary and ternary compounds in the dataset. In Fig.~\ref{fig:dataset composition}(a), which focuses on binary compounds, the majority-846 compounds (80\%)-exhibit FM ordering, while 205 compounds (20\%) are FiM. Fig.~\ref{fig:dataset composition}(b) shows the corresponding distribution for ternary compounds, where a greater number of materials are observed overall. Among these, 3,637 compounds (78\%) are FM and 1,053 compounds (22\%) are FiM.

\begin{figure}[t!] 
    \centering
    \includegraphics[width=1\linewidth]{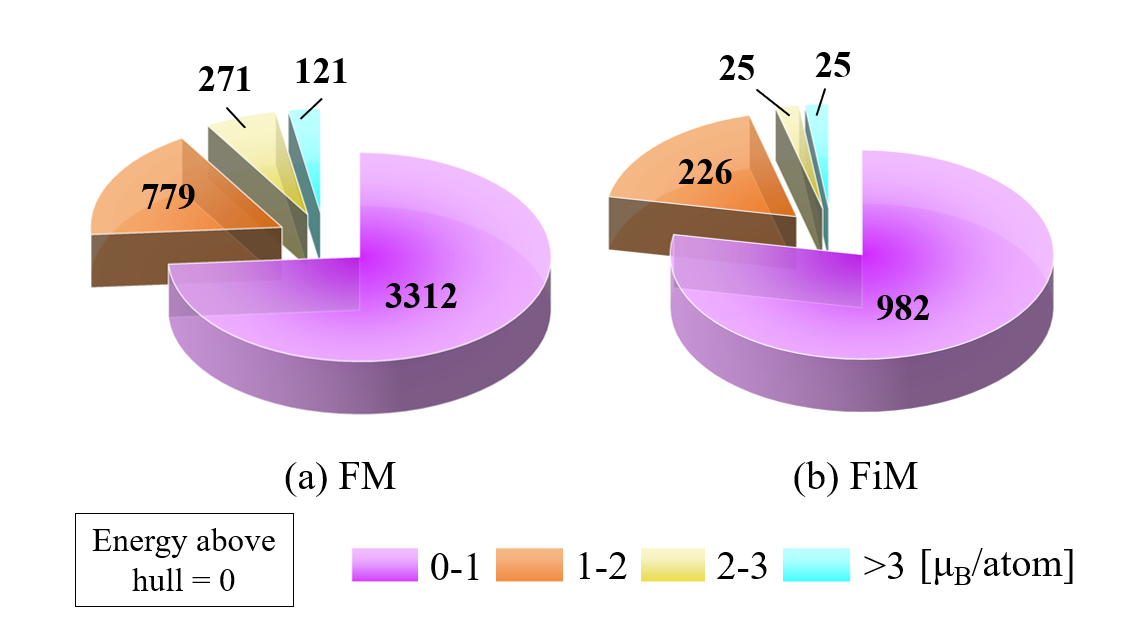}
    \caption{(a) Distribution of magnetic moments per atom for FM compounds (b) Distribution for FiM compounds. The majority of materials in both categories exhibit low magnetic moments ($<   $1 $\mu_B$/atom). A steep decline is observed with increasing magnetic moment, with only a small subset of materials exceeding 3 $\mu_B$/atom.}
    \label{fig:ub per atom} 
\end{figure}

Following the analysis of magnetic ordering, the distribution of magnetic moments per atom in the selected binary and ternary compounds is examined. This provides insight into the extent of local magnetic contributions across different compositions. The corresponding distributions are presented in Fig.~\ref{fig:ub per atom}. A detailed analysis reveals a strong prevalence of materials with low atomic magnetic moments. Specifically, 4,294 compounds - approximately 75\% of the total - exhibit magnetic moments less than 1 $\mu_B$/atom. This subset comprises 3,312 FM and 982 FiM materials. As magnetic moment per atom increases, compound numbers steeply decrease - 1,005 compounds (17.5\%) in the 1-2 $\mu_B$/atom range, 296 (5.1\%) in 2-3 $\mu_B$/atom, and only 146 (2.5\%) above 3 $\mu_B$/atom. At the extreme end of the distribution, only 13 compounds exhibit moments exceeding 5 $\mu_B$/atom. These outliers are exclusively binary systems and are predominantly composed of lanthanide elements, where the presence of highly localized 4$f$ electrons contributes significantly to large magnetic moments. Among these, europium (Eu) - based compounds account for seven, while gadolinium (Gd) - based systems constitute the remaining six.

\subsection{Feature Engineering} 

Feature engineering is the process of transforming raw data into meaningful features and is one of the persistent challenges limiting the broader success of ML in magnetic materials discovery. The effectiveness of ML models heavily depends on efficient representation of atomic and molecular structures \citep{needofdescriptors}. Descriptors encode structural and chemical information in a symmetry-invariant manner, that allows ML models to efficiently identify meaningful patterns without being affected by multiple orientations of the same material structure. Structural descriptors focus on encoding the spatial arrangement of atoms capturing features such as bond lengths, angles, and coordination numbers while composition-based descriptors represent the chemical build-up of the material, considering factors like atomic fractions, elemental types, and atomic interactions. A schematic overview of a typical ML pipeline for materials discovery, highlighting the role of descriptors, is presented in Fig.~\ref{fig:descriptor}. Various structural descriptors have been developed to encode atomic environments and interactions, including the Coulomb matrix \citep{coulomb}, Ewald sum matrix \citep{ewald}, sine matrix \citep{ewald}, many-body tensor representation (MBTR) \citep{mbtr}, atom-centered symmetry functions (ACSF) \citep{acsf}, spin-dependent ACSFs (sACSFs) \citep{sacsf}, smooth overlap of atomic positions (SOAP) \citep{soap}, and the orbital-field matrix (OFM) \citep{ofm}. These descriptors have demonstrated success in predicting key material properties such as atomization energy, formation energy, polarizability, and potential energy surfaces.  Each of these descriptors captures specific aspects of atomic and molecular systems. However, their applicability to magnetic properties remains relatively limited. Among these methods, the OFM \citep{ofm} descriptor has been applied to predict local magnetic moments in bimetal alloys with high accuracy. Nevertheless, it is predominantly constrained to ferromagnetic configurations.

The Hund's matrix method \citep{KHATRI2024172026} has been applied to only Mn-based compounds and the recall value for FiM class in FiM/FM classification is just 0.47. In this method ,the primary limitation arises from the fact that the representation of an element typically only incorporates information from the valence shell electronic configuration. To address this, we propose constructing a more comprehensive $16 \times 1$ descriptor vector for each element, which captures 8 essential ground state properties, alongside their squared values. This approach enriches the representation by incorporating both linear and nonlinear features, enabling ML models to better capture complex behaviors  \citep{Hastie2009}. The 8 properties chosen for the elemental descriptor are - atomic number (Z), atomic radius (R), group number (G), period number (P), density (D), ionization energy (I), electronegativity (E) and the number of unpaired valence electrons (U). The selection of properties used to construct a 16 $\times$ 1 vector representing an element is carefully based on their relevance to both the chemical and physical behavior of the element, providing a comprehensive understanding of its characteristics. Elemental data has been obtained from the PubChem Periodic Table provided by the National Center for Biotechnology Information \citep{PubChemPeriodicTable}. Each property captures a specific aspect of the element’s interactions and its role in various chemical processes.

Group number indicates the number of valence electrons in an element, which governs its chemical bonding behavior and typical oxidation states. Period number reflects the energy level of the valence electrons. Together, these properties encode periodic trends electronic structure. Ionization energy measures the energy required to remove an electron from an atom in its gaseous state and electronegativity quantifies an element's ability to attract electrons in a chemical bond towards itself. Both of these parameters affect bond polarity and strength of the bond formed. Elemental electronegativities can also be used to predict and understand key molecular properties such as dipole moments, bond lengths, and bond dissociation energies \citep{electronegativity}. Atomic radius influences the extent of electron cloud diffusion and nuclear attraction; as atomic radius increases, increased shielding weakens nuclear attraction to valence electrons, reducing orbital overlap and bond strength. The density of an element provides insight into its mass-to-volume ratio. Atomic number indicates the number of protons in the nucleus, strength of coulombic interactions between the nucleus and electrons and indirectly the atom's electronic configuration. This configuration directly influences the element's chemical properties, including how its electrons interact with those of other elements. A critical factor in these interactions is the number of unpaired valence electrons, which are key to an element's bonding behavior. The number of unpaired valence electrons directly impacts the element's valency, and its bonding preferences in compounds.

\begin{figure*}
    \centering
    \includegraphics[width=\linewidth]{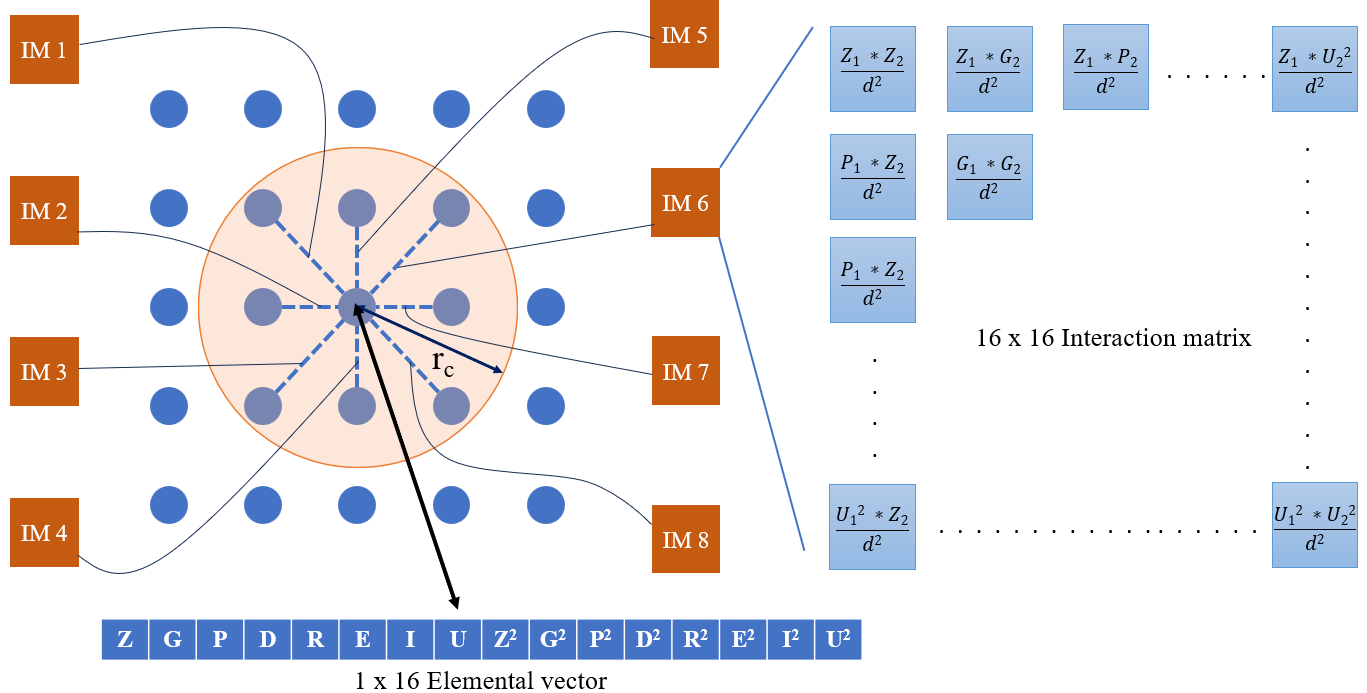}
    \caption{Illustration of the generation of a 16 $\times$ 16 interaction matrix. Each element is represented by a 16-dimensional vector incorporating atomic properties and their squared values. For each of the atoms in central unit cell, neighboring atoms within the cutoff radius are identified and interaction matrices are calculated. The matrix consists of 256 values and uniquely defines an interaction between an atomic pair.}
    \label{fig:interaction matrix} 
\end{figure*}

To construct a compound descriptor from elemental atomic descriptors, the interactions between atomic pairs is captured, while accounting for the periodicity of the lattice. Each atom in the system is represented by a 1 $\times$ 16 elemental vector, encoding its properties. The interaction between two atoms \textit{i} and \textit{j} is given by the interaction matrix (IM) of their respective elemental vectors - \( e_i \) and \( e_j \), as shown in Eq. \ref{eqn:OPM}. This IM captures the pairwise interaction between atoms and serves as the basis for constructing the compound descriptor.

\begin{equation}
\text{IM} = \frac{\mathbf{e}_i^T \times \mathbf{e}_j}{d^2}
\label{eqn:OPM}
\end{equation}

where $\mathbf{e}_i^T$ is the transpose of the vector \( e_i \) and thus IM of the elemental vectors \( e_i \) and \( e_j \) is a two-dimensional matrix as shown in Fig.~\ref{fig:interaction matrix} and each IM contains 256 values that uniquely and quantitatively describe the interaction between a specific atomic pair. The IM of each interaction has been scaled by a factor of $\frac{1}{d^2}$ (where \textit{d} is the distance between the considered atomic pair) to account for the fact the farther the atoms weaker is the interaction between them. Only atomic pairs within a specified cutoff radius \( r_c \) are considered for interaction, with \( r_c \) set to 5 \text{\AA} in this work. A range from cutoff radius from 3 \text{\AA} to 10 \text{\AA} were tested and best results were obtained at \( r_c \) set to 5 \text{\AA}. This cutoff ensures that only atoms within this distance contribute to the interaction term. To accurately model the periodic boundary conditions, interactions are computed within a 3 $\times$ 3 $\times$ 3 supercell, which includes the central unit cell and its 26 neighboring unit cells. Consequently, each atom in the central unit cell interacts not only with atoms in the same unit cell but also with atoms in adjacent cells within the supercell. Although the unit cell itself is defined with periodic boundary conditions, the model does not inherently recognize this periodicity. Therefore, the supercell is explicitly built to capture interactions across periodic boundaries. Importantly, interactions are computed only for atoms located in the central unit cell. For each of these atoms in central unit cell, neighboring atoms within the cutoff radius are identified - including those located in the surrounding (replica) cells of the supercell - but the atoms in these neighboring cells are only included if they lie within \( r_c \) of a central unit cell atom. Thus, the calculation proceeds by iterating over each atom in the central unit cell, identifying its interacting neighbors in the full supercell and constructing an interaction matrix for each interaction in cutoff radius.

Once the $16 \times 16$ interaction matrices for all atomic pairs within a compound are computed, the next step involves calculating the element-wise mean and standard deviation matrices to construct a comprehensive representation of the compound’s interactions.

First, the element-wise mean of the interaction matrices is calculated to derive a $16 \times 16$ mean matrix (\( \mathbf{M_{\text{mean}}} \)) -

\begin{equation}
\mathbf{M_{\text{mean}}} = \frac{1}{N} \sum_{k=1}^{N} \mathbf{IM_k}
\end{equation}

where \( \mathbf{IM_k} \) represents the $k$-th interaction matrix, and $N$ is the total number of interaction matrices for that particular compound. The mean matrix \( \mathbf{M_{\text{mean}}} \) represents the average interaction profile across all atomic pairs within the compound, capturing the central tendency of the pairwise interactions.

To quantify the variability of the interactions, the element-wise standard deviation matrix (\( \mathbf{M_{\text{SD}}} \)) is computed -

\begin{equation}
\mathbf{M_{\text{SD}}} = \sqrt{\frac{1}{N} \sum_{k=1}^{N} \left( \mathbf{IM_k} - \mathbf{M_{\text{mean}}} \right)^2}
\end{equation}

The standard deviation matrix \( \mathbf{M_{\text{SD}}} \) provides a measure of the dispersion or variability in the pairwise interactions, highlighting the extent to which individual atomic interactions deviate from the mean interaction profile. The resulting mean and standard deviation matrices together form the compound feature set, encapsulating both the average behavior and the variability of atomic interactions within the system. Each of these matrices have 256 values and thus in total 256 $\times$ 2 = 512 features. These matrices serve as the comprehensive descriptor for the compound, which can be utilized in ML models for predictive tasks. 

This refined descriptor offers several key advantages over Hund’s matrix. By incorporating explicit atomic properties, it ensures that chemically distinct elements receive unique representations, addressing the issue of indistinguishability in Hund’s matrix. The inclusion of squared terms enhances the ability of ML models to capture nonlinear relationships \citep{Hastie2009} between atomic properties, which can be particularly relevant in complex material behaviors. Furthermore, by reducing sparsity and increasing information density, the approach improves computational efficiency and enables more robust learning from data.

Additionally, for each atomic pairwise interaction, three key properties are captured - bond length, electronegativity difference and the difference of electronegativity squared. These features are essential for understanding the nature and strength of atomic interactions in a compound. Bond length plays a critical role in understanding the structural properties of a compound. A shorter bond length generally indicates stronger atomic interactions while longer bond lengths often suggest weaker interactions which can be indicative of less stable bonds or weaker forces. The electronegativity difference between two atoms serves as an indicator of bond polarity and bonding character. Larger differences suggest a higher degree of charge transfer and are commonly associated with ionic bonds, while smaller differences are indicative of covalent bonding, where electron sharing dominates. The difference of electronegativity squared provides a nonlinear measure of how significant the difference is between the atoms’ electronegativity values.

Once these properties are calculated for all interactions within the compound, they are converted to compound level descriptors using statistical measures - mean and standard deviation, which provide a comprehensive overview of the compound's atomic interactions. The mean values for each of these features offer a central tendency for the interactions within the compound. The mean bond length reflects the average structural characteristics of the material, while the mean electronegativity difference indicates the overall polarity or bond type (ionic or covalent). The standard deviation for each of these properties captures the variability in the atomic interactions. A higher standard deviation in bond lengths, for example, may suggest a compound with a more varied structural geometry or weaker consistency in bonding. Similarly, a high standard deviation in electronegativity differences indicates significant diversity in bond character (ranging from ionic to covalent) across the material, which can have important implications for its reactivity and bonding potential. Thus, these 6 features - mean and standard deviation of bond length, electronegativity difference and the difference of electronegativity squared - are added to the previously generated 512 features and in total 518 features are generated for a compound.

\section{Results and Discussion}

In this study, a refined descriptor for predicting the magnetic properties is proposed, constructed using a more comprehensive elemental vector representation. Unlike previous works that focus primarily on Mn-based compounds \citep{KHATRI2024172026} and bimetal alloys of lanthanide metal and transition-metal \citep{ofm}, this method is applied to a more generalized dataset comprising 5741 stable, binary and ternary, FM and FiM compounds available in the Materials Project database \citep{Jain2020TheMP}. This diverse dataset encompasses a broader spectrum of chemical compositions and structural motifs compared to datasets that utilize elemental or symmetry filters. The inclusion of a wide variety of compositions and structures introduces additional variability, making the task of modeling magnetic behavior more intricate and requiring advanced approaches to capture the underlying trends and interactions.

The model is trained to predict two important magnetic properties solely from the structural information of the compound - magnetic ordering - FM or FiM and magnetic moment per atom ($\mu_{\text{B}}$/atom). For this task, LightGBM \citep{lightgbm}, a gradient boosting framework known for its efficiency and performance with large datasets, is employed. The model's accuracy is measured using four goodness-of-fit metrics - the coefficient of determination ($R^2$), correlation coefficient (CC), mean absolute error (MAE), and root mean square error (RMSE), comparing predicted values to the true data. $R^2$ measures the proportion of variance in the true data explained by the model and is given by -

\begin{equation}
R^2 = 1 - \frac{\sum_{i=1}^{n}(y_i - \hat{y}_i)^2}{\sum_{i=1}^{n}(y_i - \bar{y})^2}
\end{equation}

where $\bar{y}$ is the mean of the observed values. An $R^2$ value of 1 indicates perfect predictions, while 0 implies that the model does no better than predicting the mean of the observed values. Negative values suggest that the model performs worse than a constant mean predictor. 

CC evaluates the linear relationship between predicted and actual values -
\begin{equation}
CC = \frac{\sum_{i=1}^{n}(y_i - \bar{y})(\hat{y}_i - \bar{\hat{y}})}{\sqrt{\sum_{i=1}^{n}(y_i - \bar{y})^2} \sqrt{\sum_{i=1}^{n}(\hat{y}_i - \bar{\hat{y}})^2}}
\end{equation}

where $\bar{y}$ and $\bar{\hat{y}}$ are the means of the actual and predicted values, respectively. The value of $CC$ ranges from -1 to 1, where 1 indicates a perfect positive linear relationship, 0 indicates no linear correlation, and -1 indicates a perfect negative linear relationship.

MAE represents the average magnitude of prediction errors -
\begin{equation}
MAE = \frac{1}{n} \sum_{i=1}^{n} |y_i - \hat{y}_i|
\end{equation}

It is a linear score, meaning all individual differences are weighted equally in the average. Lower values of MAE imply more accurate predictions.

RMSE is similar to MAE but penalizes larger errors more heavily -
\begin{equation}
RMSE = \sqrt{\frac{1}{n} \sum_{i=1}^{n} (y_i - \hat{y}_i)^2}
\end{equation}

This metric penalizes larger errors more heavily than MAE, making it sensitive to outliers. A thorough understanding of magnetic ordering-whether it be FM, FiM, AFM, or other complex forms of magnetic order (helimagnets, spin glasses, skyrmion lattices) - remains crucial for the development of materials with specific magnetic characteristics. The arrangement of magnetic moments within a material determines its overall magnetic behavior, influencing phenomena such as hysteresis, coercivity, and susceptibility. Magnetic ordering also governs the response of the material under various thermal and external field conditions, which is of particular importance for the stability and performance of devices. Equally important to the overall magnetic behavior is the magnetic moment per atom - a fundamental intrinsic property that governs a material’s macroscopic magnetic behavior and plays a pivotal role in its technological applications. Magnetic moment not only influences the material's response to external magnetic fields but also affects its interactions with other magnetic entities at the atomic and molecular levels. Different applications often demand materials with specific ranges of magnetic moments. The accurate prediction of both of these properties of a material is not only critical for the rational design of new materials but also for the optimization of existing devices that rely on these magnetic properties.

\subsection{Magnetic ordering prediction}

\begin{figure}[t!] 
    \centering
    \includegraphics[width=0.8\linewidth]{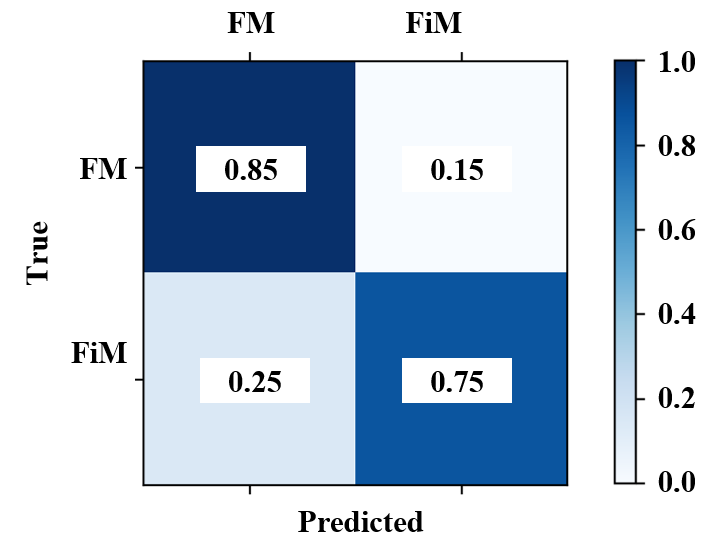}
    \caption{Confusion matrix for FM vs. FiM magnetic ordering classification using the proposed descriptor. It highlights the model’s balanced class-wise accuracy in predicting magnetic ordering, with a significant enhancement in correctly identifying FiM compounds (recall= 0.75) compared to the Hund’s matrix approach (recall = 0.47 for FiM)}
    \label{fig:confusion matrix} 
\end{figure}

From ML standpoint, distinguishing between FM and FiM compounds is more difficult than separating FM/FiM from AFM ones, due to the overlapping global magnetic features shared by FM and FiM systems - both exhibit non-zero net magnetization and undergo similar thermal behavior -specifically, Curie-like magnetic phase transitions. This overlap leads to feature space entanglement, making it harder for ML models to learn discriminative patterns, whereas AFM compounds form a more separable class due to their distinct zero net magnetization and Néel-type transitions, facilitating clearer decision boundaries. The model achieved an impressive overall accuracy of 82.4\% in predicting the magnetic ordering, a substantial improvement compared to the 72.88\% accuracy reported by previous models using Hund’s Matrix \citep{KHATRI2024172026}. An accuracy of 80.5\% was achieved using random forest regressor. The classification between FM or FiM materials is critical in real-world applications because both types are used in permanent magnets, but they exhibit distinct behaviors that influence their performance. FM materials have parallel aligned atomic moments, resulting in strong and uniform magnetization, making them ideal for applications requiring stable and high magnetic fields, such as motors, transformers, and magnetic data storage. On the other hand, FiM materials have antiparallel atomic moments of unequal magnitude leading to a lower net magnetization and lower coercivity. While this results in a material with a weaker overall magnetic response compared to FM, this same feature provides distinct advantages in certain applications. The lower coercivity in FiM materials means they are easier to magnetize and demagnetize, making them suitable for applications where a more reversible, dynamic magnetic response is needed. For instance, this characteristic is beneficial in magnetic memory devices, where fast switching between magnetic states is essential for efficient data storage and retrieval. Therefore, accurately distinguishing between FM and FiM materials enables the targeted design of magnets with specific properties suited to industrial applications and technological advancements. 

One of the key strengths of the model is its balanced performance across both magnetic ordering classes. While the Hund’s Matrix model showed a recall of 0.86 for FM compounds and a recall of only 0.47 for FiM compounds, the present model delivers a recall of 0.85 for FM compounds and a much improved recall of 0.75 for FiM compounds. This enhanced recall for FiM compounds is particularly noteworthy, given that FiM materials constitute the minority class (22\% of total) in the dataset making them prone to being underrepresented during training and are characterized by complex and subtle magnetic interactions. The confusion matrix for the proposed descriptor is shown in Fig.~\ref{fig:confusion matrix}. The model’s refined ability to handle these complex interactions underscores its potential for advancing material design in magnetic technologies.

\subsection{Magnetic moment prediction} 

\begin{table}[t!]
\centering
\caption{Test set performance metrics of the model for $\mu_{\text{B}}$/atom prediction}
\label{table:m_prediction}
\renewcommand{\arraystretch}{1.5}  
\begin{tabular}{|>{\centering\arraybackslash}p{3.5cm}|>{\centering\arraybackslash}p{1.5cm}|>{\centering\arraybackslash}p{1.5cm}|>{\centering\arraybackslash}p{1.5cm}|}
\hline
\textbf{Descriptor} & \textbf{CC} & \textbf{MAE} & \textbf{RMSE} \\ \hline
Hund's matrix \citep{KHATRI2024172026}       & 0.6885      & 0.2051        & 0.31    \\ 
Orbital field matrix \citep{ofm}                 & 0.7138      & 0.1979       & 0.30     \\ 
This work     & 0.9395      & 0.1862        & 0.28            \\ \hline
\end{tabular}
\end{table}

\begin{figure}[t!] 
    \centering
    \includegraphics[width=0.9\linewidth]{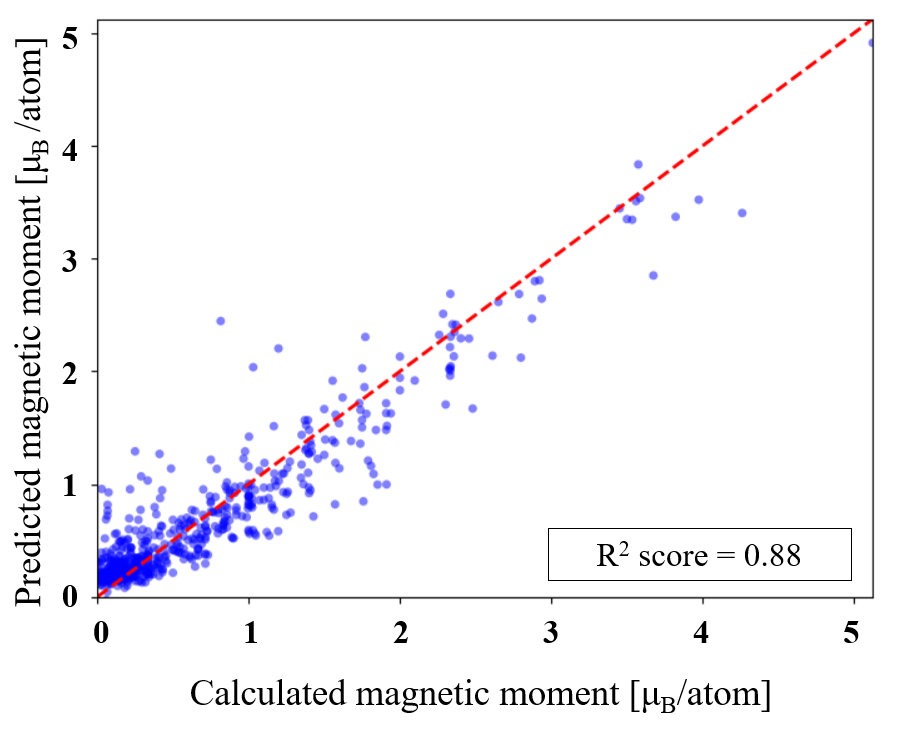}
    \caption{The plot shows the predicted magnetic moment per atom (using the refined descriptor) versus the actual values from the test set.}
    \label{fig:m prediction} 
\end{figure}

The proposed method represents a significant advancement in the accuracy of magnetic moment predictions, achieving an impressive CC of 0.93 for the magnetic moment per atom. CC of 0.89 was achieved using random forest regressor. This notable improvement surpasses the performance of both the Hund’s matrix \citep{KHATRI2024172026} descriptor and the Orbital field matrix \citep{ofm}, which are commonly used in previous models. The enhanced predictive capability of this method is attributed to its ability to capture more intricate atomic-level interactions and dependencies, leading to more reliable estimations of magnetic properties. The model parameters - CC, MAE and RMSE - on the test set are shown in Table ~\ref{table:m_prediction}. Fig.~\ref{fig:m prediction} presents a comparison between the predicted magnetic moment per atom and the actual values for the test set, utilizing the refined descriptor. The strong correlation between the predicted and actual values is clearly evident, with most data points clustering closely around the red dashed \textit{y} = \textit{x} line. This alignment indicates a high degree of accuracy in the model’s predictions, demonstrating that the refined descriptor effectively captures the underlying relationships between atomic characteristics and magnetic moments. The improved accuracy can be attributed to three major factors - (1) the incorporation of a broader set of elemental properties (2) the extension of the descriptor to include both linear and nonlinear features (by incorporating squared terms), and (3) the reduction in sparsity of matrix by encoding detailed atomic information. Additionally, training on a diverse dataset of stable binary and ternary compounds improves generalizability, enabling accurate predictions across varied chemical environments and magnetic orders. The model serves as a robust tool for materials discovery, offering high predictive performance while maintaining computational efficiency.

\subsection{Formation energy prediction}

\begin{figure}[t!] 
    \centering
    \includegraphics[width=0.9\linewidth]{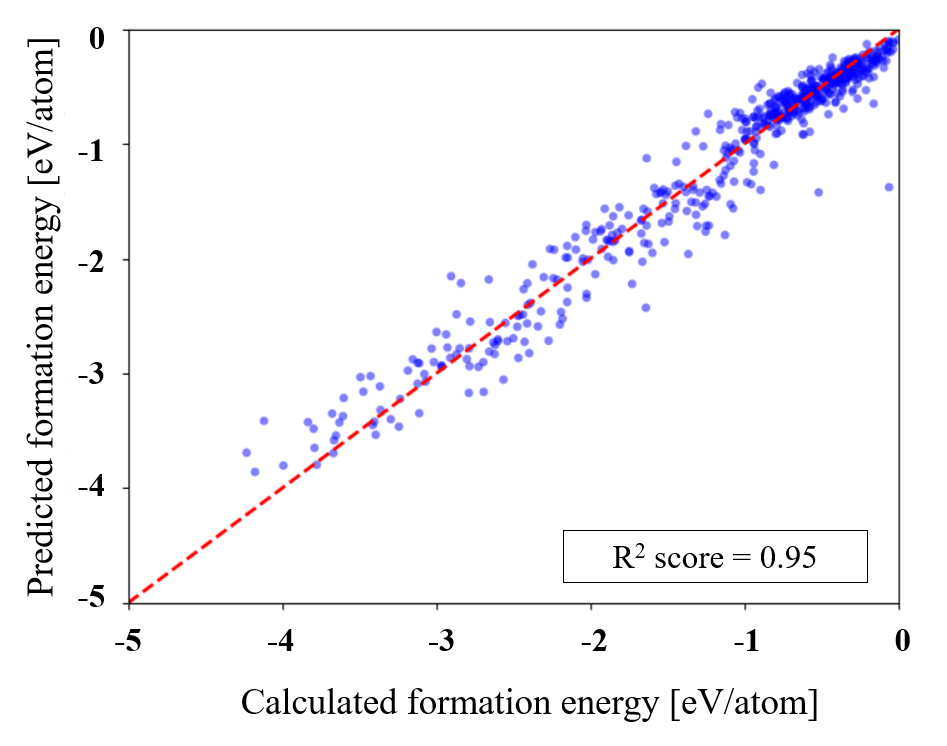}
    \caption{The plot shows the predicted formation energy per atom (using the refined descriptor) versus the actual values from the test set.}
    \label{fig:fe prediction} 
\end{figure}

The model is also trained to predict formation energy per atom and a CC of 0.97 is attained. CC of 0.94 was achieved using random forest regressor.The prediction of formation energy per atom is important as it reflects the stability of a material and its synthesizability, helping to identify promising candidates for practical use. The model parameters on the test set are shown in Table ~\ref{table:fe_prediction}. Fig.~\ref{fig:fe prediction} shows the predicted magnetic moment per atom versus the actual values for the test set using our refined descriptor. The strong correlation, with most points clustering around the red dashed \textit{y} = \textit{x} line, highlights the model’s accuracy.

\begin{table}[t!]
\centering
\caption{Test set performance metrics of the model for formation energy/atom prediction}
\label{table:fe_prediction}
\renewcommand{\arraystretch}{1.5}  
\begin{tabular}{|>{\centering\arraybackslash}p{3.5cm}|>{\centering\arraybackslash}p{1.5cm}|>{\centering\arraybackslash}p{1.5cm}|>{\centering\arraybackslash}p{1.5cm}|}
\hline
\textbf{Descriptor} & \textbf{CC} & \textbf{MAE} & \textbf{RMSE} \\ \hline
Hund's matrix \citep{KHATRI2024172026}       & 0.946              & 0.189       & 0.335     \\ 
Orbital field matrix \citep{ofm}                 & 0.961             & 0.157       & 0.285     \\ 
This work     & 0.977           & 0.144      & 0.203      \\ \hline
\end{tabular}
\end{table}

\section{Conclusion}

Proposed descriptor achieves a high correlation coefficient of 0.93 for predicting magnetic moment per atom, significantly outperforming previous methods such as the Hund’s matrix \citep{KHATRI2024172026} and Orbital field matrix \citep{ofm}. The model achieves 82.4\% accuracy in predicting magnetic ordering (FM vs FiM), outperforming previous methods and offering much improved recall for FiM ordering. Enhanced performance is driven by the - (1) Inclusion of a comprehensive set of elemental properties, (2) The extension to capture both linear and nonlinear features through squared terms and (3) A reduction in sparsity by incorporating detailed atomic interactions within the matrix. The model’s robustness is further enhanced by training on a diverse dataset of stable binary and ternary compounds, ensuring its generalizability across various chemical environments and magnetic ordering types. This approach demonstrates strong predictive capability while maintaining computational efficiency, positioning it as a powerful tool for materials discovery, particularly in the search for novel magnetic materials with optimized properties.

\section{Acknowledgments}

The high performance computational facilities viz. Aron (AbCMS lab, IITB), Dendrite (MEMS dept., IITB), Spacetime, IITB and CDAC Pune (Param Yuva-II) are acknowledged for providing the computational hours. JJ acknowledges funding received through PMRF grant (PMRF ID : 1300138). AB thanks funding received through BRNS regular grant (BRNS/37098) and SERB power grant (SPG/2021/003874).

\bibliography{prop}

\bibliographystyle{apsrev4-2} 


\end{document}